# Nearly quantum limited nanoSQUIDs based on cross-type Nb/AlO$_x$/Nb junctions

M Schmelz[1*], V Zakosarenko[2], T Schönau[1], S Anders[1], S Linzen[1], R Stolz[1], and H-G Meyer[1]

[1] Leibniz Institute of Photonic Technology, Albert-Einstein-Straße 9, D-07745 Jena, Germany
[2] Supracon AG, An der Lehmgrube 11, D-07751 Jena, Germany

* Corresponding author. Tel.: +49 3641 206122; fax: +49 3641 206199
  E-mail address: matthias.schmelz@ipht-jena.de

**Abstract**
We report on the development of nearly quantum limited SQUIDs with miniature pickup loop dimensions. The implemented high quality and low capacitance cross-type Nb/AlO$_x$/Nb Josephson junctions offer large $I_CR_N$-products and therefore enable an exceptional low noise level of the SQUIDs. Devices with loop dimensions of 1 µm exhibit white flux noise levels as low as 45 nΦ$_0$/Hz$^{1/2}$ corresponding to an energy resolution ε of about 1 $h$ at 4.2 K, with $h$ being Planck's constant.
Moreover, the large usable voltage swings of the devices of about 300 µV allow highly sensitive and easy single-stage operation while exploring nearly the intrinsic noise of the SQUIDs, beneficial e.g. for sensor arrays in SQUID microscopy.

Keywords: nanoSQUIDs, cross-type Josephson junction, flux noise, quantum limit

## 1. Introduction

Magnetic imaging [1] or the investigation of magnetic nanoparticles [2], molecular magnets [3] as well as single electrons or cold atoms [4] are emerging fields in e.g. modern material science. In recent years, several magnetic imaging tools emerged, like e.g. magnetic resonance force sensors [5], sensors based on nitrogen vacancy defects in diamond [6], and Hall [7] or SQUID type sensors [8]. Although the latter are one of the most sensitive devices for measuring magnetic flux, their usual optimization with respect of field sensitivity or current resolution leads to devices with dimensions of several µm to mm and are therefore not well adapted to the task of measuring small spin systems.

For nanoscale magnetic sensing the figure of merit is the spin sensitivity $S_\mu^{1/2} = S_\Phi^{1/2}/\Phi_\mu$ with $S_\mu$ and $S_\Phi$ being the noise spectral density normalized to magnetic moment and flux, respectively, and $\Phi_\mu$ the coupling between a particle with magnetic moment $\mu$ to the SQUID. Downsizing the SQUID loop dimensions improves $S_\mu^{1/2}$ as it reduces the equivalent flux noise spectral density $S_\Phi$ via the decrease in total SQUID inductance $L_{SQ}$ [9] and increases $\Phi_\mu$ as well [10, 11].

Therefore, during the last years several micrometer or even nanometer-sized SQUID sensors have been developed. A recent review about the various approaches can be found in [12] and will also be given within this edition.

Most of the recent approaches are based on constriction type junctions, fabricated e.g. by patterning a small hole in a planar thin superconducting strip either by electron beam lithography or focused ion beam etching [13-15]. By doing so, white flux noise levels of down to 0.2 µΦ$_0$/Hz$^{1/2}$ have been achieved [13]. As the critical current of the constriction is temperature dependent, these sensors, however, have only a limited temperature working range for optimum performance. Moreover, for optimum coupling the magnetic particle has to be placed close to the constriction, preventing the independent optimization of device performance and coupling $\Phi_\mu$.

In order to potentially reduce the distance between the sensor and the sample (and therefore inevitable losses), superconducting thin-films have as well been deposited on non-planar surfaces [16]. There, the SQUID loop is formed on the apex of a hollow quartz tube pulled into a very sharp pipette resulting in white flux noise levels of down to 50 n$\Phi_0$/Hz$^{1/2}$ for Pb based devices [17]. Recent activities with additional Josephson junctions to extend the working point with optimal sensitivity [18] show very promising results. However, such Pb based devices are known to suffer from poor long term stability mainly against thermal cycling.

By implementing Nb/HfTi/Nb sandwich-type junctions together with e-beam lithography the reliable fabrication of SQUIDs with very small pickup loop areas based on tunnel junctions became possible [19] and flux noise levels of about 115 n$\Phi_0$/Hz$^{1/2}$ have been reported [20]. This technology, moreover, even allows non-planar SQUID loop geometries [21]. Beside the large critical current noise of the used SNS junctions, which leads to poor low-frequency noise of the SQUIDs, the performance of these devices probably suffer from a lower characteristic voltage than those based on comparable SIS Josephson junctions.

We therefore recently introduced a technology for the fabrication of miniature SQUIDs based on cross-type SIS Nb/AlO$_x$/Nb Josephson tunnel junctions [22]. In [23] we reported on the development of highly sensitive SQUIDs with loop dimensions ranging from 10 µm down to 500 nm. Downsizing the SQUID loop improves the flux noise spectral density $S_\Phi$ but reduces the screening parameter $\beta_L = 2L_{SQ}I_C/\Phi_0$ to values far away from the optimum $\beta_L \approx 1$ [9]. Although the SQUIDs already exhibited very low white flux noise levels below 100 n$\Phi_0$/Hz$^{1/2}$, sensors with $\beta_L \ll 1$ showed only moderate noise improvements compared to somewhat larger devices.

In this paper we will thus focus on further noise optimization of such sensors via an increase of the screening parameter $\beta_L$. In section 2 we briefly describe the sample fabrication. Device characterization and discussion of experimental data will be given in section 3. Finally we will illustrate how further optimization in terms of spin sensitivity $S_\mu^{1/2}$ can be achieved.

## 2. Sample fabrication

The devices described in this paper have been fabricated in the IPHT cross-type Josephson junction technology [22]. Originally aimed for the development of integrated SQUID sensors, it comprises three superconducting layers, several isolation layers of thermally evaporated SiO$_x$, an AuPd resistive layer and finally a thermally evaporated SiO$_x$ protection layer. It has been successfully proven for e.g. the fabrication of highly sensitive SQUID current sensors [24], SQUID magnetometer used in geophysical measurement systems [25-27] or even more complex SQUID structures with up to about 500 Josephson junctions per chip.

The integration of nanoSQUIDs within this technology offers the possibility to fabricate SQUID-arrays as well as to include adapted SQUID amplifier on the same chip, e.g. for sophisticated multiplexing readout schemes. The technology allows as well the fabrication of non-planar devices, with pickup loop geometries perpendicular to the chip surface.

For the development of planar-type highly sensitive miniature SQUIDs described within this paper, however, we use only some layers of this process. The fabrication starts with the deposition of the Nb/AlO$_x$/Nb trilayer, which is subsequently patterned into a narrow strip, forming part of the SQUID (see part A in figure 1). Figure 1 shows a scanning electron microscope image of a SQUID with inner loop diameter $w = 3.0$ µm and $(1.5 \times 1.5)$ µm$^2$ Josephson junctions. After a planarization process with thermally evaporated SiO$_x$, a Nb wiring layer is deposited on top and patterned in an "u"-shape to close the SQUID loop (part B in figure 1). This step also defines the Josephson junctions, which are formed by the overlap between these two layers and are indicated by the two dotted squares in figure 1. Due to this self-aligned process there is no undesired overlap between superconducting layers around the Josephson junctions and consequently no parasitic capacitance in parallel to the junction capacitance is formed, which results in a very low total junction capacitance of about 65 fF/µm$^2$ [28]. For the used Josephson junctions with a side length of 1.5 µm this results in a total junction capacitance of about 140 fF. An AuPd layer is used to electrically shunt the Josephson junctions.

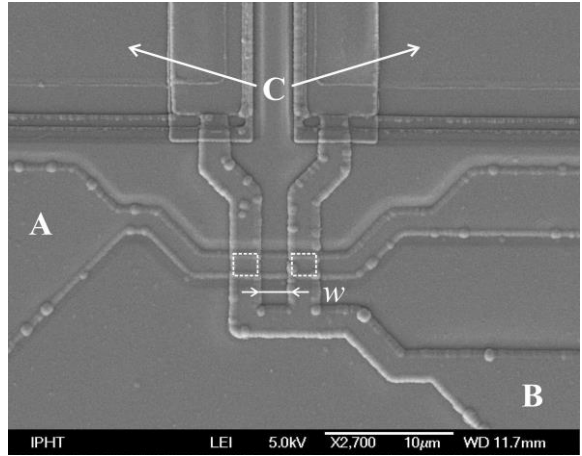

**Figure 1.** Scanning electron microscope images of a SQUID with $(1.5 \times 1.5)$ µm$^2$ Josephson junctions and an inner loop dimension $w = 3.0$ µm. The Josephson junctions (dashed area) are defined by the overlap between Nb/AlO$_x$/Nb trilayer (A) and Nb wiring layer (B). Shunt resistors (C) are used to overdamp the junctions.

### 3. Sensor Design and Device Characterization

In this paper we will mainly focus on further noise optimization of SQUIDs with small pickup loop dimensions via an increase of the screening parameter $\beta_L = 2L_{SQ}I_C/\Phi_0$. As the SQUID inductance $L_{SQ} = L_{geo} + L_{kin}$ is given by the geometrical inductance $L_{geo}$ of the SQUID loop and the kinetic inductance $L_{kin} = \mu_0 \lambda_L^2 \cdot s/(dh)$ [29] via the material and thickness/ shape of the thin-films, $\beta_L$ can be adjusted by the critical current $I_C$ of the Josephson junctions. Here, $s$ is the circumference, $d$ and $h$ are the width and height of the cross section of the SQUID loop, respectively, $\lambda_L$ is the London penetration depth and $\mu_0$ the vacuum permeability. Besides utilizing a larger critical current density, for the work described in this paper we will mostly make use of somewhat larger junction dimensions, not to impair the low-frequency noise performance of the SQUIDs. As the white flux noise of a SQUID $S_\Phi^{1/2} = 4\, L_{SQ}^{3/4}\, C_{JJ}^{1/4}\, (2k_B T)^{1/2}$ [30] depends only weakly on the junction capacitance, the drawback of larger $C_{JJ}$ is overcome by the increase in $\beta_L$.

For the devices described in this paper, we used a critical current density of about 2.5 kA/cm$^2$. This results in a critical current of about 110 µA for SQUIDs with a junction size of $(1.5 \times 1.5)$ µm$^2$. The characterization of current voltage characteristics of unshunted Josephson junctions fabricated on the same wafer indicate their high quality with ratios of the subgap to normal state resistance of about (30-40), typical for our fabrication process [22, 31]. Table I lists representative parameters of the investigated SQUIDs. We fabricated devices with loop dimensions ranging from 5 µm down to 1 µm. Smaller loop dimensions may further improve the white noise performance of devices, but have not been fabricated due to the resolution limit of the used i-line stepper lithography tool used for fast turn-around times. For these optimized devices the McCumber parameter is $\beta_C = 2\pi I_C R^2 C/\Phi_0 \approx 1.1$, which is close to the desired theoretical value for optimum noise performance [9]. Here $I_C \approx 55$ μA and $R \approx 6.6$ Ω denote the measured critical current and normal resistance of one Josephson junction at 4.2 K, and $C_{JJ} \approx 140$ fF is the total junction capacitance.

**Table 1.** Device parameters of the investigated SQUIDs: $w$ denotes the inner diameter of the SQUID loop, as indicated in figure 1 and $V_{pp}$ the usable voltage swing of the SQUID as measured at 4.2 K. The geometrical inductance has been simulated using FastHenry [32]; values of the kinetic inductance are based on $L_{kin} = \mu_0 \lambda_L^2 \cdot s/(dh)$, with $\lambda_L = 90$ nm, as given in the text. The flux noise was measured using a two-stage configuration with a SQUID current sensor as a low-noise preamplifier.

| SQUID # | loop diameter $w$ | inverse effective area [$\mu T/\Phi_0$] | $V_{pp}$ [$\mu$V] | geometrical inductance $L_{geo}$ [pH] | kinetic inductance $L_{kin}$ [pH] | $\beta_L$ | measured white flux noise [n$\Phi_0$/Hz$^{1/2}$] |
|---|---|---|---|---|---|---|---|
| 1 | 5.0 μm | 47.5 | 345 | 17.1 | 1.8 | 1.00 | 150 |
| 2 | 3.0 μm | 104 | 340 | 10.7 | 1.4 | 0.64 | 70 |
| 3 | 1.0 μm | 430 | 320 | 5.00 | 0.94 | 0.32 | 45 |

Device characterization has been carried out in a specially designed dipstick immersed in liquid helium at 4.2 K. Here, a superconducting solenoid made from NbTi wire was used to provide the magnetic background fields to set up the working points and to measure the effective areas of the SQUIDs, as listed in table I. They can be described by the simple assumption of $A_{eff} \approx w' \cdot (w' + 2 \cdot a)$, where $w' = w + 160$ nm represents the shrink during fabrication compared to the design value. Here $a$ is the linear dimension of the Josephson junction.

Figure 2 shows a representative set of voltage-flux characteristics and a current-voltage characteristic of SQUID #1. The usable voltage swing, the maximum voltage swing of the SQUID in current bias mode, accounts to about 300 μV for all investigated devices, as given in table I. Owing to the large transfer functions of up to 22 mV/$\Phi_0$ in a single stage configuration, nearly the intrinsic noise of the SQUIDs (typically within a factor of two) could be measured with state-of-the-art low noise SQUID electronics [33]. Thus, it offers highly sensitive and easy single-stage operation, beneficial e.g. for sensor arrays in SQUID microscopy. It is also worth to note that the values above indicate representative ones, e.g. the measured effective areas differ less than 5% between up to ten devices of each kind, thus demonstrating the capability of the presented approach for the reproducible fabrication of sensor arrays.

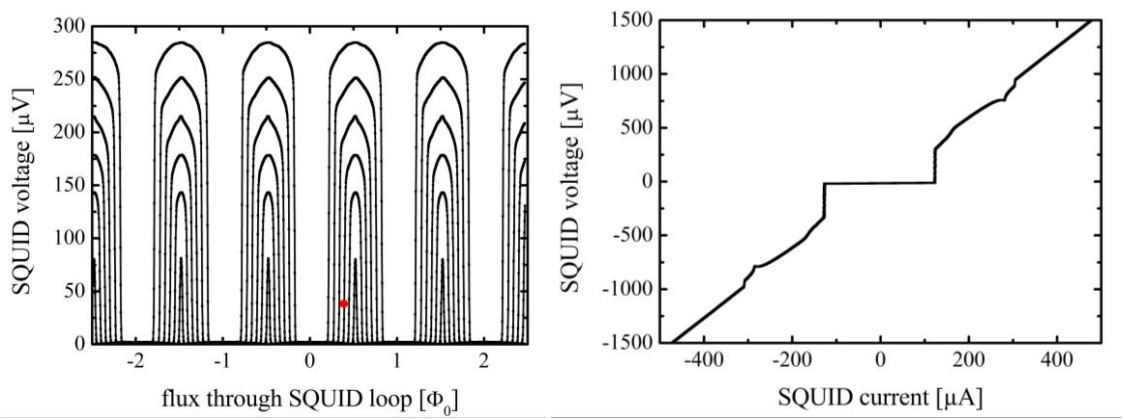

**Figure 2.** Typical set of voltage-flux characteristics (left) and current-voltage characteristic (right) of a SQUID #1 with (1.5 × 1.5) μm² Josephson junctions and an inner loop dimension $w$ of 5.0 μm for bias currents between 50 and 110 μA in steps of 5 μA. The flux was applied by a superconducting solenoid. The working point where noise measurements were performed is marked in red.

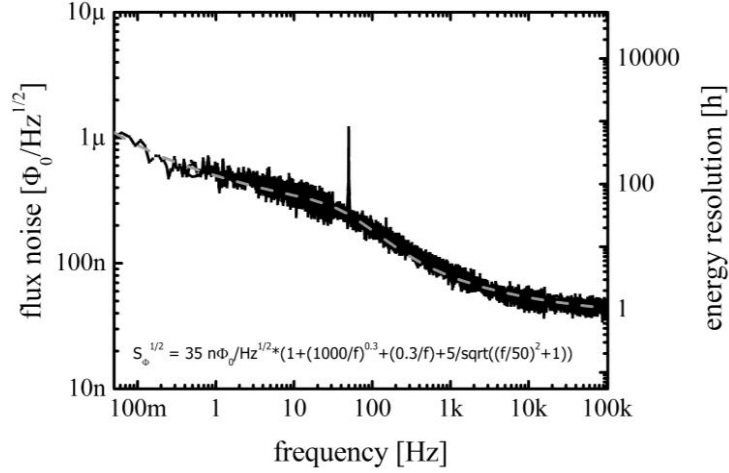

**Figure 3.** Equivalent flux noise spectrum for SQUID #3 measured at 4.2 K (black) and the fitted spectral dependence as a dashed gray line. The right hand axis has been calculated according to $\varepsilon = S_\Phi/(2 \cdot L_{SQ})$, with $L_{SQ} = 5.94$ pH as listed in table I. Note that the energy resolution in the white range accounts to about 1.1 $h$, with $h$ being Planck's constant.

To exploit the intrinsic noise performance of the SQUIDs, a two-stage setup has been used for the noise measurements. Here, a SQUID array composed of 16 SQUIDs acts as a low noise preamplifier. The SQUID to be measured was voltage biased with $R_S = 0.5\ \Omega$ and the critical current change due to an external signal is sensed with the SQUID array as an ammeter, exhibiting an input current noise of below 2.5 pA/Hz$^{1/2}$. The working point has been adjusted by the superconducting solenoid such that the signal amplitude sensed in the SQUID array due to a small magnetic field modulation to the miniature SQUID was maximum. Both SQUIDs have been placed in a superconducting and $\mu$-metal shield, and measurements have been carried out in liquid helium at 4.2 K. A commercial low-noise directly-coupled flux locked loop electronics [33] has been used for the measurements to provide the bias currents and feedback to the amplifier SQUID.

The output voltage has been recorded with an HP 3565 spectrum analyzer with a maximum bandwidth of 100 kHz. The equivalent flux noise, as shown in figure 3 for SQUID #3, results from the measured transfer function $V_\Phi$ in the corresponding working point according to $S_\Phi^{1/2} = S_V^{1/2}/V_\Phi$. The measured white flux noise levels of the investigated devices are listed in table I. For SQUID #3 the measured white flux noise is about 45 n$\Phi_0$/Hz$^{1/2}$, which corresponds to an energy resolution $\varepsilon = S_\Phi/(2L_{SQ}) = 1.1\ h$, with $L_{SQ} = 5.94$ pH. The spectral dependence can be described with $S_\Phi^{1/2} = 35$ n$\Phi_0$/Hz$^{1/2} \cdot [1+(1\text{ kHz}/f)^{0.3} + (0.3\text{ Hz}/f) + 5/\sqrt{((f/50\text{ Hz})^2 + 1)}]$, as shown by a dashed gray line in figure 3. Please note that according to this fit, even at 100 kHz there may be a fraction of "low-frequency noise" adding to the intrinsic white noise of the SQUID. The measured flux noise at 1 Hz amounts to about 0.53 $\mu\Phi_0$/Hz$^{1/2}$ which is equivalent to $\varepsilon \approx 150\ h$, a typical value for SQUIDs fabricated within this technology.

Accordingly, there are different noise sources present in our devices. The first is white noise, whose magnitude will be discussed below. Second, the noise in the frequency range between 1 Hz and about 10 kHz typically shows a dependence of $S_\Phi^{1/2}$ proportional to $1/f^\alpha$, with $\alpha \approx 0.3$-0.5. This contribution vanishes in magnetic insensitive working points with $\partial V/\partial \Phi = 0$, which rules out critical current fluctuations. Instead it is related to a magnetic signal and has as well been reported in other SQUIDs [25, 34-36]. Since flux trapping in the SQUID washer can be neglected due to the narrow linewidth of the superconductor [28, 37, 38], it is very unlikely that motion of magnetic vortices trapped in the SQUID washer causes this noise. Up to now there is no comprehensive understanding of this noise source and we currently attribute this to an unknown effect such as fluctuation of surface spins e.g. at the superconductors surface [39]. Below about 0.3 Hz the noise is dominated by critical current fluctuations in the Josephson junctions. If the empirical formula describing the $1/f$ critical current fluctuations in $AlO_x$ based Josephson junctions [40] is used, a magnitude of critical current fluctuations of about 138 pA/Hz$^{1/2}$ at T = 4.2 K and $f$ = 1 Hz for SQUID #3 can be estimated. Taking into account a typical current sensitivity of LTS SQUIDs of $M_{dyn} = (\partial\Phi/\partial I) \approx L_{SQ}$ [41], this results in a magnitude of flux noise at 1 Hz of about 0.4 $\mu\Phi_0$/Hz$^{1/2}$ which is close to the measured value. In addition, there is a kink visible in the noise spectrum at about 10 Hz, which can be described by a low-pass filter behavior with a cut-off frequency of about 50 Hz. It is due to a magnetic signal, probably caused by eddy currents nearby the SQUID loop or a nearby single fluctuator, which however needs further study.

In order to compare the measured white flux noise levels to theoretical predictions, we will follow up the discussion presented in [23]. Therein, the widely known relation for the energy resolution of a SQUID $\varepsilon = 16 k_B T (L_{SQ} C_{JJ})^{1/2}$ is considered as the thermal energy $k_B T$ distributed in a frequency range limited by the SQUID time-constant $\tau_{LC} = (L_{SQ} C_{JJ})^{1/2}$, which holds for the condition $\beta_C \approx \beta_L \approx 1$. As this may not be true for all of our devices, we substitute the $LC$ time-constant with an effective time-constant $\tau_{eff}$ given as $1/\tau_{eff} = 1/\tau_{LC} + 1/\tau_{RC} + 1/\tau_{LR}$, with $\tau_{RC} = RC$ and $\tau_{LR} = L_{SQ}/R$, resulting in $\varepsilon = 16\sqrt{2} k_B T \tau_{eff}$ for the energy resolution and correspondingly $S_\Phi^{1/2} = (32\sqrt{2} L_{SQ} k_B T \tau_{eff})^{1/2}$ for the expected white flux noise of the device. Using this equation, the measured white flux noise levels are in good agreement with theoretical predictions and typically show only approximately 30% higher values.

It is furthermore worth to discuss the fundamental noise limitation of the SQUID: In [17] the quantum noise limit is given as $S_\Phi^{1/2} = (\hbar L_{SQ})^{1/2}$, with $\hbar = h/2\pi$ being the reduced Planck's constant. For SQUID #3 this results in $S_\Phi^{1/2} = 12.1$ n$\Phi_0$/Hz$^{1/2}$. The measured equivalent white flux noise levels are therefore only a factor of about four larger than the quantum noise limit.

Finally we estimate the spin sensitivity of these devices. According to [11] the coupling of a point-like dipole with magnetic moment of one Bohr magneton $\mu_B$ in the center of a square SQUID washer with inner side length $2a = 1.0$ µm is given as $\Phi_\mu = \sqrt{2}/\pi \cdot \mu_0 \mu_B/a = 5.1$ n$\Phi_0/\mu_B$. Thus the measured flux noise of SQUID #3 in the white noise region of 45 n$\Phi_0$/Hz$^{1/2}$ results in a white-noise spin sensitivity of about 9 $\mu_B$/Hz$^{1/2}$.

## 4. Conclusion

In conclusion, we reported on the development of nearly quantum limited SQUIDs featuring small pickup loop dimensions. Due to the high quality and low capacitance cross-type Josephson junctions, they exhibit very low white flux noise levels down to 45 n$\Phi_0$/Hz$^{1/2}$. For devices with 1 µm loop diameter the energy resolution thus corresponds to 1.1 h at 100 kHz.

The large usable voltage swings of about 300 µV offer highly sensitive and easy single-stage operation while exploring almost the intrinsic noise of the SQUIDs. Together with the reproducible fabrication of such devices within this sophisticated technology this allows for the development of sensor arrays that offer advantages for imaging methods such as e.g. SQUID microscopy.

The individual noise sources in the investigated devices have been revealed and described, which may allow for further noise optimization especially in the low-frequency range. Further downsizing the SQUID loop dimension will presumably not only further improve the white noise performance but will moreover lead to enhanced coupling to a weak magnetic moment. Beyond that, the presented devices allow for the independent optimization in terms of noise and coupling, e.g. due to the use of local constrictions in the SQUID loop, potentially leading to single spin resolution.